\documentstyle{ptptex}

\notypesetlogo  

\preprintnumber{
NSF-ITP-95-XXX\\hep-th/9511157}

\title{%
Recent Results in String Duality
}

\author{%
Joseph {\sc Polchinski}\footnote{E-mail address: joep@itp.ucsb.edu}
}

\inst{%
Institute for Theoretical Physics
\\
University of California
\\
Santa Barbara, CA 93106-4030
}

\recdate{%
\today
}

\abst{%
This is a talk given at YKIS '95, primarily to non-string theorists.
I review the evidence for string duality, the principle that any string
theory at strong coupling looks like another string theory at
weak coupling.
A postscript summarizes developments since the conference.
}

\begin{document}

\maketitle

When I gave this talk in late August I resolved to write it up
immediately, because the field was moving very fast and the
`recent' results would quickly become dated.  Well, as I sit down to
write it is actually two and a half months later, and indeed many
things have happened since my talk.  So I will write up the talk
as given, and add a postscript to describe a few of
the developments since.

\section{Introduction}

Until recently, one might have expected that string theories at
strong coupling would involve new and exotic physics.  For example,
since string theory includes gravity, one might have expected
a phase with large fluctuations of the spacetime geometry.  But
now it appears that string theory at strong coupling is not so
exotic.  Rather, {\it string duality} is the principle that any string
theory at strong coupling simply looks like another string theory at
weak coupling.

This sort of duality is familiar in low dimensional field theories.
That it might happen in string theory has been
put forward by a number of people over the years, and
was extensively pursued by Duff, Sen, Schwarz, and others.
Until recently though, it seemed unlikely to many of us that
such a thing could happen even in a nontrivial four dimensional field
theory, and much more unlikely that it could happen in a theory as
complicated as string theory.  But the work of Seiberg and others
has made it clear that this can indeed happen in four-dimensional
gauge theories, and now it appears overwhelmingly likely that it
happens in string theory as well.

String duality is a revolution, shaking our understanding
of the foundations of the theory.  Many of the
world-sheet properties that previously received great emphasis now
appear to be technical features of string perturbation theory, not
preserved by duality.
On the other hand,
spacetime ideas such as supergravity are playing a more prominent
role in determining the structure of the theory.
At this point it is not clear where we are headed or whether the final
result will still be called `string theory.'
The main theme of my talk is ``Should you believe it?''
Along the way I will try to discuss a few of the open questions and
some of what has been learned.

Here is a list of a few of the conjectured dualities, in various
numbers of dimensions:
\begin{eqnarray}
d=10:\quad && \mbox{IIA string}\ \leftrightarrow\ \mbox{`M-theory'
on $S_1$}\nonumber\\
&& \mbox{$SO(32)$ heterotic string}\ \leftrightarrow\
\mbox{$SO(32)$ type I string}\nonumber\\[4pt]
d=7:\quad && \mbox{heterotic string on $T^3$}\
\leftrightarrow\ \mbox{M-theory on $K3$}\nonumber\\[4pt]
d=6:\quad && \mbox{heterotic string on $T^4$}\ \leftrightarrow\ \mbox{IIA
string on $K3$}\nonumber\\[4pt]
d=4:\quad && \mbox{heterotic string on $T^6$}\ \leftrightarrow
\ \mbox{heterotic string on $T^6$} \label{duals}
\end{eqnarray}
This is a remarkable list: on the one hand it looks quite
complicated---in each of the listed dimensions the heterotic string is
dual to a different theory.  But the structure is quite constrained,
and fits together in an intricate way as one compactifies and
decompactifies dimensions.
By the way, all the theories on this list have a large amount of
supersymmetry, the equivalent of $N=4$ in four dimensions.  For
theories with less supersymmetry the dynamics and dualities are even
richer.

A skeptic might take the point of view that all evidence for string
duality is circumstantial, and that the successes of string duality
are consequences only of the strong constraints that supersymmetry
imposes on the low energy field theory.  A severe skeptic might make
this same argument in regard to weak/strong duality in $N=4$
supersymmetric {\it field} theory, while a less severe one might
believe that duality is possible in field theory but does not extend
to the full string spectrum.
Both of these points of view were reasonable at one time (or at
least I hope they were, because I held them), but the evidence has
rapidly mounted, to the point that the issue has shifted from ``is it
true?'' to ``what does it mean?''

In this talk I will try to assemble the evidence.  This will be far
from a systematic review of the subject, but rather a presentation
of those issues to which I have given the most thought.

\section{Evidence for String Duality}

\subsection{Heterotic $S$-Duality in $d=4$.}

A great deal of attention has been given to the heterotic string
compactified on a 6-torus.  This is supposed to have a infinite
discrete symmetry $SL(2,Z)$, generated by a weak/strong
duality transformation combined with discrete
shifts of the $\theta$-parameter\cite{font}.  The evidence, reviewed in
ref.~\citen{sen}, is very similar to that for the older conjecture
of duality in $N=4$ supersymmetric field theory: invariance of the
BPS mass formula for stable electric and magnetic charges, and
of the lattice of charges allowed by the Dirac
quantization condition.  This is not convincing to either skeptic: it could
just be a consequence of supersymmetry.  It would be less trivial,
and say more about the dynamics, if one could show that the
{\it actual} spectrum of BPS states (the degeneracies of the allowed
states) is dual.  There is some evidence here, the existence of a
monopole bound state required by string duality.  But this is
only evidence for duality in {\it field} theory.  A search for the duals of
string states has been inconclusive, depending on an understanding of
soliton collective coordinates at the string scale.\cite{Hmon}

\subsection{Heterotic String on $T^4$ $\leftrightarrow$
IIA String on $K3$}

In six dimensions, the heterotic string on a 4-torus and the Type IIA
string on $K3$ (the only 4-dimensional Calabi-Yau manifold) have
the same low energy field theory and in fact the same space of
vacua, the coset space
\begin{equation}
O(\Lambda_{4,20})\backslash O(20,4;R)/O(20,R) \times O(4,R).
\end{equation}
Low energy supersymmetry by itself requires that the space locally
be of the form
\begin{equation}
O(n,4;R)/O(n,R) \times O(4,R).
\end{equation}
So the fact that $n=20$ in both cases is a nontrivial
coincidence.\cite{seiberg}  That the global structures (the
discrete identifications) match is a further nontrivial fact,
involving stringy states and not just the low energy spectrum: the
discrete identification on the heterotic side includes $T$-duality,
interchanging winding and momentum states, while the discrete
identification on the IIA side includes mirror symmetry and discrete
shifts of world-sheet $\theta$-parameters.\cite{AM}

A possible contradiction arises because the IIA string has an Abelian
gauge group~$U(1)^{24}$, while the heterotic string has this group
in generic vacua but has unbroken non-Abelian symmetries at special
points.  The mechanism to resolve this is the same as that found by
Strominger for the conifold~\cite{strom} (discussed further below): when a
nontrivial surface in the compact dimensions shrinks to zero size, a
wrapped soliton can become massless, providing the needed gauge bosons.
Indeed, Witten~\cite{wit1} argued that the $K3$ theories with enhanced
gauge symmetry contained collapsed spheres.  There was still a puzzle
because these $K3$'s are orbifolds and so correspond to solvable and
nonsingular conformal field theories, where string perturbation
theory should be a good qualitative guide.  The existence of a
massless soliton, a large nonperturbative effect, would then be
surprising and disturbing.  It would mean that perturbation
theory can break down without warning, a severe
problem since we have no other definition of string theory.  Again
this is evaded: Aspinwall showed that the orbifold and
enhanced-symmetry theories differ by a background antisymmetric
tensor field, so it is quite possible that the latter theory
corresponds to a singular CFT as is the case for the conifold.

\subsection{The Big Picture}

In his famous talk at USC,\cite{wit1} Witten proposed that every
string theory in every dimension has a strong coupling dual.\footnote
{To be precise, the conjecture holds in this simple form only for
theories with at least $N=4$ supersymmetry, where the BPS formula
allows a global definition of the dilaton.}  Collecting
some earlier duality conjectures, discarding others, and adding
some of his own, he produced a nearly complete
set of duals (much of this list was anticipated in ref.~\citen{HT}).
A skeptic could argue that most of this paper is based on low energy
field theory and on unproven conjectures about the existence of BPS
states.  But there are a many nontrivial checks in it.  In
particular, although the pattern of duals has an intricate
dependence on dimension (see the earlier list) compactification never
leads to two {\it different} weakly coupled dual candidates for a
given theory, which would be a contradiction.  There is always
exactly one candidate.  Moreover, for all the dual pairs, the field
redefinition that relates the low energy theories always includes a
sign change of the dilaton (else one would have a weak/weak duality
of different string theories, again a contradiction).

\subsection{The Conifold Transition}

For several years there has been a puzzle that certain Calabi-Yau
compactifications are singular,\cite{coni} with various couplings
diverging.  Although these couplings are calculated at string
tree level, for the Type II string there are nonrenormalization
theorems so that in some cases (the vector multiplet Kahler
potential) the result is known to be free of perturbative {\it and
nonperturbative} corrections.\cite{strom}  That is, the exact
effective action is singular.  This was shown to have a simple and
natural interpretation in terms of a massless soliton (see above).
But in some cases\cite{GMS} there is a new branch of vacua in which
the soliton has an expectation value.  This highly nonperturbative
phase has a natural dual description in terms of a weakly coupled
string theory on a Calabi-Yau space of different topology.

Incidentally, an important open question is to understand the precise
nature of the breakdown of perturbation theory at the conifold and
enhanced gauge symmetry points.  There are interesting conjectures
here\cite{wit2,matmod} but no clear picture.

\subsection{Heterotic String as a Soliton}

In addition to light states from small loops, string
theories include strings of macroscopic size.  An infinite
straight string (or a string wound around a large periodic dimension)
is a BPS state, invariant under half the supersymmetry, and so is stable
under changes in the parameters.  Taking the example of the heterotic
string theory in six dimensions, start with a
macroscopic heterotic string in the weakly coupled theory.  Increase the
coupling until the strong coupling limit, described by the weakly coupled
IIA string, is reached.  A state which looks like a long heterotic string
must still exist but is no longer present as a fundamental string state:
it must be a soliton.

Indeed, the IIA theory contains a soliton with
exactly the properties of the heterotic string.\cite{hetsol1}
In particular, the degrees of freedom of the soliton are precisely those
of the fundamental string:\cite{hetsol2} transverse oscillations moving on
the torus of the dual heterotic theory, right-moving spinors and a
left-moving $E(8) \times E(8)$ current algebra.  This is a necessary check
on string duality, and is a strong piece of positive evidence.  Start from
the weakly coupled IIA theory, in which this heterotic soliton is heavy.
As the coupling is increased the BPS formula implies that the soliton
becomes lighter, and in the strong coupling limit its tension is the
smaller than any other scale: it is hard to see how the effective theory
could then be anything but the heterotic string.\footnote
{It should be noted that the converse check, finding the IIA string as a
soliton in the heterotic theory, is less clear-cut.  A soliton carrying
the appropriate charge exists---this by itself is rather trivial---but its
degrees of freedom are harder to determine, as they involve subtleties
in the quantization of solitons below the string scale.}

\subsection{Loop and Nonperturbative Corrections}

Direct checks of weak/strong duality are difficult because one can only
calculate in the weakly coupled theory.  But in theories with at least
$N=2$ supersymmetry, there are nonrenormalization theorems which guarantee
that some amplitudes calculated at weak coupling are in fact exact in the
quantum theory.  The strategy is a common one in string theory: the
coupling, and therefore the quantum corrections, depend on the dilaton
and supersymmetry restricts the way the dilaton can appear in the
effective Lagrangian.  With $N \geq 4$ supersymmetry there are a few
couplings that can be compared in this way,\cite{rad1} but the case $N=2$
is particularly rich: an infinite number of couplings (an entire function)
can be compared; for a review see ref.~\citen{rad2}.  This appears to be
a very nontrivial check, relating string tree and loop calculations, as
well as nonperturbative results, on the two sides.

An important open question is to understand the strong-coupling behavior
of $N=2$, 1, and~0 theories as completely as that for $N=4$.
Their richer dynamics makes these theories more interesting (and of course
we live in an $N=0$ or approximately $N=1$ vacuum), but also makes it
harder to solve them.  Another key question is how string
duality relates to all the recent results on supersymmetric gauge
theories.  Obviously there must be a close connection, but as yet the
two subjects are surprisingly disjoint.  For example, in the gauge
theories the focus is on the physics at long distance, whereas string
duality appears to hold at all scales.

\subsection{CHL Models}

The final check I will describe in more detail, not because it is
especially important but because it is the one in which I have personally
been involved, and serves to illustrate some important ideas.  Let me
first recall the beautiful work of Narain,\cite{narain} who described the
space of vacua of the toroidally compactified heterotic string.  This
displays several important phenomena in string theory:
the existence of
a moduli space of degenerate but physically inequivalent vacua,
a discrete group of equivalences (dualities),
points of enhanced gauge symmetry, and limits (decompactifications)
in which the theory goes over to seemingly different ten-dimensional
string theories, the $E(8) \times E(8)$ and $SO(32)$ heterotic strings.
Note that all of this is perturbative, holding in the weakly coupled
theory---it is not the nonperturbative string duality that is currently
causing so much excitement.  But in fact a central theme of string duality
is that more of these same phenomena arise nonperturbatively in the larger
moduli space that includes the string coupling (dilaton).  In fact, the
dilaton, which plays such an important role in string perturbation theory,
is not distinguished in the full quantum theory: it is on an equal footing
with the moduli from the compactification.

Strong coupling duals for these $N=4$ theories were
proposed in refs.~\citen{sen,HT,wit1}.  It was widely assumed that
these were the only $N=4$ vacua of the heterotic string,\footnote
{Except for some unpublished work of Dixon, quoted in the
review~\citen{schell}.} but
Chaudhuri, Hockney, and Lykken (CHL)~\cite{CHL} pointed out the existence of
many new $N=4$ vacua.  It turns out that all of these can be identified with
toroidal compactifications of the heterotic string,\cite{CHL1,CHL2} but
with fields periodic only up to a discrete global symmetry such as the
interchange of the two $E(8)$'s of the heterotic string; Narain's work
had included periodicity up to a {\it local} symmetry.

Considering first $d=4$, the Narain compactifications are supposed to be
self-dual under weak/strong duality ($S$-duality).\cite{font,sen}  For the
CHL theories there is a complication.  Narain compactification gives
simply-laced gauge groups ($SU(n)$, $SO(2n)$, $E(n)$), but the CHL
theories include non-simply laced groups.  The moduli space was explored
in ref.~\citen{CHL2} and was found to include points with symmetries
\begin{equation}
Sp(20) \times SO(17-2d), \quad Sp(18) \times SO(19-2d), \quad \ldots
\ , \quad Sp(2d) \times SO(37-4d). \label{groups}
\end{equation}
Now, $S$-duality includes electric/magnetic duality in the low energy
theory.  This takes simply-laced groups into themselves, but interchanges
long and short roots and so takes $Sp(2n) \leftrightarrow
SO(2n+1)$.\cite{GNO}  For Narain compactifications, $S$-duality acts
trivially on the moduli, but in the CHL theories it must act on the moduli
so as to move the theory from a point of one symmetry to a point of dual
symmetry.  Now there is a nontrivial check, because gauge groups had
better appear in the moduli space in dual pairs.  This need only hold in
$d=4$, because only in this case is there electric/magnetic duality: only
in four dimensions do the field strength $F_{\mu\nu}$ and its dual have
the same rank (this also gives some idea as to why the pattern of
string dualities depends so strongly on dimension).  Examining the
list~(\ref{groups}), we see that for general $d$ the set is not dual, but
precisely in $d=4$ it is.  Nothing in the compactification distinguishes
$d=4$, and duality of the groups is not automatic in some trivial way, but
precisely in $d=4$ where it must appear it does.

One might wonder whether this is an independent check.  In fact it can
almost be derived from facts we already know, but the derivation is a nice
illustration of how the whole pattern of string dualities fits together.
To start, let us ask how it is that the toroidally compactified heterotic
string is dual to itself in $d=4$ but to the IIA string in $d=6$: how
do these fit together if we compactify the $d=6$ theory on $T^2$?
The answer is that the group of dualities is actually very large, with a
given string theory generally having many self-dualities (both
perturbative and non-perturbative) as well as equivalences to other
string theories.  But at a given point in parameter space at most one
of this set will have a weak coupling and a natural spacetime
interpretation,\footnote{The latter requirement excludes compactification
radii shorter than the string scale and so factors out the perturbative
$T$-dualities.} which is one of the checks mentioned before.  Which theory
is weakly coupled is determined by $d$-dependent dimensional
analysis,\cite{wit1} leading to the intricate pattern~(\ref{duals}).

If two string theories are equivalent, their self-duality groups must be
conjugate: `duality of dualities.'\cite{dudu}  In particular, the
nonperturbative $S$-duality of the $d=4$ heterotic string is conjugate to
the perturbative $T$-duality of the IIA string on $K3 \times T_4$:
\begin{equation}
S = \beta T \beta^{-1}
\end{equation}
where $\beta$ stands for the $d=6$ string-string duality.  The
non-perturbative $S$-duality of the $d=4$ heterotic string thus follows
from the $\beta$ symmetry.  The same is in principle true of the
$S$-duality of the CHL theories.\cite{wv1}\footnote{
Ref.~\citen{wv1} does not find the full $S$-duality group, in particular
not the transformations which act nontrivially on the moduli.  These must
involve a product of $T$ and a symmetry acting on the $K3$
moduli space.}

In $d=6$ the CHL theories present a puzzle: here the Narain
theories are dual to the IIA theory on $K3$, but there is no other $N=4$
IIA compactification in conformal field theory.  This was elegantly resolved
in ref.~\citen{ss}, which proposed a dual with a nontrivial Ramond-Ramond
background field.  Such a background cannot be described in conformal
field theory and is not well-understood as yet.  The arguments in
ref.~\citen{ss} use another strategy common in string theory.  Start
from the $d=11$ M-theory compactified on a product $K3 \times S_1$.
Consider compactification first on $K3$ and then on $S_1$, and then in
the reverse order.  Using the duality conjectures~(\ref{duals}) for
compactification of the M-theory, one obtains a known dual pair, the
heterotic theory on $T^4$ and the IIA theory on $K3$.  Now adding a
twist to the compactification (identifying under a combined shift on the
$S_1$ and isomorphism on the $K3$), one obtains the dual pair of
ref.~\citen{ss}, one of which is the $d=6$ CHL theory.

\section{Discussion}

The picture that emerges from all of this is of a single theory, with a
space of vacua labeled by the dilaton and the moduli from the
compactification. In the bulk of moduli space the theory is fully quantum
mechanical, but it has many classical limits.  The asymptotics in the
various classical limits are given by string perturbation theory, and all
of the seemingly different supersymmetric string theories appear as limits
of this single theory.

Coming back to the question ``Should you believe it?'' and speaking as a
former skeptic, I would say that my skepticism ended around section~2.5,
the heterotic string as a soliton.  Too many things work.  Test after
test that could have failed succeeds, to the point that the simplest
explanation is that all of these theories are connected.

The question now is, ``What is the theory?''  Is it even a string theory?
One point of view would be that all of the various string theories can be
formulated nonperturbatively as string field theories, and then
transformed into one another by an appropriate change of variables.  This
is parallel to what is possible in the Thirring/Sine-Gordon duality, the
prototype of a quantum-mechanical equivalence.  However, I think that the
evidence points in another direction, namely that the various string
theories are just generators of the asymptotics of the theory at weak
coupling and below the Planck scale.  Even before string duality there were
reasons to believe this, such as the limited success of string field
theory, and the rapid growth of string perturbation theory.\cite{shenk}
In the matrix model for example, the rapid growth of string perturbation
theory is a sign that the string are best thought of as a
composite,\cite{LH} in that case of free fermions.
String duality gives a further insight: strings appear in the various limits
simply because the BPS formula requires them to be the lightest objects in
the theory, and not necessarily because they play a fundamental role.

In some ways the situation is much like the 60's and early 70's, with a
great deal of `data' and no theory.  We know a lot about the
asymptotics of the theory, and a bit about how they fit together.  An
interesting clue is the strong coupling limit of the $d=10$ IIA string,
which appears to be an eleven-dimensional theory.  There is no perturbative
string theory in eleven dimensions.  Evidently the long distance physics is
$d=11$ supergravity, but the short distance physics is unknown and for
now is referred to as `M-theory.'  This theory has no coupling constant
(dilaton) and so is intrinsically quantum mechanical.

Does string duality help address the phenomenological problems
of string theory, the cosmological constant and the choice of vacuum?
As yet the results are disappointing, in that there is no new physics
at strong coupling but just more of what has been seen at weak
coupling.\footnote{See however the final section of ref.~\citen{wit2}.} But
it is not time to be impatient.  We are learning
remarkable and surprising things about string theory.  That test after
test of string duality is working tells us that there is a new structure
to be found.  Once it is found we may hope that we will learn new things
about the dynamics of string theory and the structure of the vacuum.
Indeed, it would be disappointing if we were able to find the right string
`model' without first answering the question, ``What is string theory?''

\section{Postscript}
The field has continued to move at a rapid rate.  We still have not
answered the question posed at the end of the previous section, but the
web of connections between the different string theories has gotten much
tighter.  Some of the developments:

\subsection{D-Branes}

In closed string theories, a string state can be roughly factored into the
separate states of the right-moving and left-moving degrees of freedom.
The type II string thus has two kinds of boson, the
Neveu-Schwarz/Neveu-Schwarz (NS\,NS) states which are products of bosons,
and the Ramond/Ramond (RR) states which are products of fermions.  Both
sectors include gauge fields, and in fact the RR sector includes
generalized gauge fields, antisymmetric tensors of various ranks.
Fundamental strings carry only the NS\,NS charges, but string duality
requires states in the spectrum which carry RR charge as well.
Previously~\cite{HT,wit1,strom} the necessary states were described as
black holes.  This was not entirely satisfying, however, because of the
singularity and the difficulty of quantization.

It was argued some time ago~\cite{DLP} that in addition to fundamental
string states, string theory necessarily contained `D-branes,' extended
objects swept out by the endpoints of open strings.  It turns
out\cite{dbrane} that these objects also carry the RR charge.  The quantum
is precisely that required by the Dirac quantization condition, and by
string duality.  Thus it seems that these are the RR-charged objects
need for string duality.  This is a more precise description than
the black hole, at least for states of small charge.  Such a simple
description is apparently possible because the RR-charged objects are
lighter than ordinary solitons (one less power of the coupling) and so are
in some sense a smaller disturbance (analogous to the single-eigenvalue
tunneling in the matrix model).

The identification of the RR solitons as
D-branes has made suddenly possible many explicit studies of the spectrum,
in agreement with duality.\cite{braneaps}  At the same time, however, it
points up even more than before that we are only working with an
effective theory.  A sum over all virtual D-branes would be
extremely unwieldy and is unlikely to be correct.  Rather, the D-brane
description is likely valid only at scales long compared to that set by
the mass (or tension) of the D-brane, so such a sum is
inappropriate.\footnote
{Except in the special circumstance that a D-brane is light due to a
degenerating cycle.}
There is an interesting question as to the physical size of D-brane,
and of the fundamental degrees of freedom that we are looking for.  Shenker
has suggested that there is evidence for a scale shorter than the string
scale.\cite{shenk2}  This is at first sight plausible, as the Planck
scale now seems to play a more central role than the string scale.  But I do
not know of any sense in which the effective size of the D-brane is
smaller than the string scale.  See refs.~\citen{Dsize} for further
investigation of this.

\subsection{Type I--Heterotic Duality}

Of the various dualities conjectured in ref.~\citen{wit1}, the $d=10$
$SO(32)$ type I--heterotic duality was somewhat separate from the others,
and had the least supporting evidence.  Subsequently a heterotic soliton
was found in the type I theory, but it is of a particularly singular sort
so one cannot be sure that it is in the spectrum.\cite{huldab}  Indeed,
the strongest argument for this duality was sinply ``What else?''---given
the evidence in other cases that string duality is a general principle,
this is the natural pairing to make in $d=10$.

There is now new evidence.\cite{polwit}
First, following the idea of `duality of dualities,' we can ask how the
$T$-dualities of the compactified heterotic string map to the type I
theory.  The answer is that they imply a rather complicated self-duality
of the type I theory, one which does not hold in perturbation theory.
This is an apparent contradiction, because the type I theory seems to be
weakly coupled in the relevant region.  But a more careful examination
shows that perturbation theory breaks down in a novel and intricate way,
at the precise point in parameter space where string duality requires
light nonperturbative states to appear.  Second, the heterotic soliton
in the type I theory carries RR charge.  As noted above, it now appears
that such solitons should be described as D-branes.  The relevant D-brane
has precisely the world-sheet structure of heterotic string.

\subsection{The $E(8) \times E(8)$ Theory}

A notable gap in ref.~\citen{wit1} was the absence of a candidate dual
for the $d=10$ $E(8)\times E(8)$ heterotic string.  This gap has now been
filled.  Compactifying the $d=11$ M-theory on a circle
gives the IIA string.  Ref.~\citen{horwit} elegantly argues that
compactifying the same theory on a line segment produces the $d=10$
$E(8)\times E(8)$ heterotic string, with the gauge symmetry living on
the boundaries of spacetime.  This is further evidence for the relevance
of the $d=11$ theory as well.

\subsection{Conclusion}

Again and again the predictions of string duality are borne out, often in
surprising ways.  It appears that at present the theory is smarter than we
are, knowing how to connect many disjoint pieces of physics and
mathematics.  It remains to unravel the hidden structure that makes this
possible.  It is notable that many interesting but separate lines of
development from recent years---mirror symmetry, supergravity,
supersymmetric solitons, D-branes, and others---have now come together to
play key roles in string duality.  It is likely that there are other past
developments whose significance has not yet been realized, and which will
in their turn help to complete the picture.

\section*{Acknowledgements}
I would like to thank S. Chaudhuri, A. Strominger, and E. Witten for
enjoyable collaborations, and in particular A. S. for helping me to catch
up with this subject.  I would also like to thank the many ITP postdocs
and visitors with whom I have discussed these issues.

\end{document}